\documentclass[11pt,fleqn,twoside]{article}
\usepackage{amsfonts,amssymb,latexsym}
\makeatletter
\newcommand{\prava}[1]{\small\it
\begin{flushleft}
Copyright \copyright \ 1999 by  #1
\end{flushleft}}

\newcommand{\name}[1]{\begin{flushleft}
                       \LARGE \bf #1
                       \end{flushleft}\vspace{-3mm}}

\newcommand{\Author}[1]{\begin{flushleft}
                       \it #1 \end{flushleft}}

\newcommand{\Adress}[1]{\begin{flushleft}
                       \it #1 \end{flushleft}}

\newcommand{\Date}[1]{\begin{flushleft}
                      \small  \it #1 \end{flushleft}}

\newcommand{\ehkol}{Author \ name}
\newcommand{\ohkol}{Article \ name}
\renewcommand{\@evenhead}{
\hspace*{-3pt}\raisebox{-15pt}[\headheight][0pt]{\vbox{\hbox to \textwidth 
{\thepage \hfil \ehkol}\vskip4pt \hrule}}}
\renewcommand{\@oddhead}{
\hspace*{-3pt}\raisebox{-15pt}[\headheight][0pt]{\vbox{\hbox to \textwidth 
{\ohkol \hfil \thepage}\vskip4pt\hrule}}}
\renewcommand{\@evenfoot}{}
\renewcommand{\@oddfoot}{}

\newcommand{\be}{\begin{equation}}
\newcommand{\ee}{\end{equation}}
\newcommand{\ba}{\hspace*{-5pt}\begin{array}}
\newcommand{\ea}{\end{array}}

\makeatother

\newtheorem{theorem}{Theorem}
\newtheorem{lemma}[theorem]{Lemma}
\newtheorem{prop}[theorem]{Proposition}
\newcommand{\til}{\tilde}
 
 \newcommand{\dl}{\delta}

\newcommand{\rh}{\rho}  
\newcommand{\ph}{\phi} 
\newcommand{\phv}{\varphi} \newcommand{\ch}{\chi}
 \newcommand{\ps}{\psi} 
\newcommand{\om}{\omega} \newcommand{\Om}{\Omega}
 \newcommand{\raw}{\rightarrow}
 
\newcommand{\g}{\frak g} \newcommand{\F}{\frak F}
\newcommand{\C}{{\mathbb C}} 
\newcommand{\bib}{\bibitem} 
 
\renewcommand{\H}{\mbox{$\cal H$}}  
\newcommand{\n}{\parallel} \newcommand{\h}{{\bf h}}
 \renewcommand{\ll}{\label}
 
\newcommand{\R}{{\mathbb R}} 
  \newcommand{\notp}{p \kern-.48em /} \newcommand{\ci}{\cite}
\newcommand{\bea}{\begin{eqnarray}} \newcommand{\eea}{\end{eqnarray}}
\newcommand{\ot}{\otimes}
\newcommand{\half}{\mbox{\footnotesize $\frac{1}{2}$}}
\newcommand{\ca}{$C^*$-algebra}
\newcommand{\jlt}{\tilde{J_L}}
\newcommand{\enp}{\hfill $\blacksquare$}
\renewcommand{\F}{{\cal F}}

 \newcommand{\wmn}{({\sf m,n})}
\newcommand{\pmn}{U^{({\sf m,n})}} \newcommand{\pim}{U^{{\sf m}}}
\newcommand{\pin}{\overline{U}^{{\sf n}}}
\newcommand{\Hmn}{{\cal H}^{({\sf m,n})}}
\newcommand{\Hm}{{\cal H}^{{\sf m}}}
\newcommand{\Hn}{\overline{{\cal H}}^{{\sf n}}}
\newcommand{\plm}{U_{{\sf m}}} \newcommand{\pll}{U_{{\sf l}}}
\newcommand{\plla}{U_{{\sf l}'}}

\newcommand{\Hlmn}{{\cal H}_{({\sf m,n})}}
\newcommand{\Hlm}{{\cal H}_{{\sf m}}}
\newcommand{\Hll}{{\cal H}_{{\sf l}}}
\newcommand{\Hul}{{\cal H}^{{\sf l}}}

\newcommand{\lu}{{\mathbf u}_0({\cal H})^*}
\newcommand{\Omn}{{\cal O}_{\sf m,n}}
\newcommand{\pq}{\pi_{\rm qua}}

\newcommand{\pl}{\hbar} 
\newcommand{\rip}{(\cdot,\cdot)_0} \newcommand{\pco}{U_{\rm co}}
\newcommand{\m}{{\bf m}} 

\newcommand{\res}{\upharpoonright} \newcommand{\plc}{U_{\chi}}
\newcommand{\puc}{U^{\chi}} \newcommand{\hlc}{{\cal H}_{\chi}}
 \newcommand{\huc}{{\cal H}^{\chi}}
\newcommand{\la}{\langle} \newcommand{\ra}{\rangle}
\renewcommand{\g}{{\mathbf g}} \newcommand{\rep}{representation}
\newcommand{\irep}{irreducible representation}

\newcommand{\MW}{Marsden-Weinstein}

\renewcommand{\O}{{\cal O}}
\newcommand{\PH}{{\mathbb P}{\cal H}}
\newcommand{\I}{{\mathbb I}}
\newcommand{\Lk}{\overline{\cal L}_k}

\begin{document}
\thispagestyle{empty}
\setcounter{page}{161}

\renewcommand{\ehkol}{N.P. Landsman}
\renewcommand{\ohkol}{Representations of the Inf\/inite Unitary
Group from Constrained Quantization}

\begin{flushleft}
\footnotesize \sf
Journal of Nonlinear Mathematical Physics \qquad 1999, V.6, N~2,
\pageref{landsman-fp}--\pageref{landsman-lp}.
\hfill {\sc Article}
\end{flushleft}

\vspace{-5mm}

\renewcommand{\footnoterule}{}
{\renewcommand{\thefootnote}{} \footnote{\prava{N.P. Landsman}}}

\name{Representations of the Inf\/inite Unitary
Group from Constrained Quantization}\label{landsman-fp}

\renewcommand{\thefootnote}{$^*$}

\Author{N.P. LANDSMAN~\footnote{Supported by a fellowship from the Royal
Netherlands Academy of Arts and Sciences (KNAW)}}

\Adress{Korteweg-de Vries Institute for Mathematics, University of
 Amsterdam\\
 Plantage Muidergracht 24, 1018 TV Amsterdam, The Netherlands \\[1mm]
 E-mail: npl@wins.uva.nl}

\Date{Received September 18, 1998; Accepted December 1, 1998}

\begin{abstract}
\noindent
We attempt to reconstruct the irreducible unitary \rep s of the Banach
Lie group $U_0(\H)$ of all unitary operators $U$ on a separable
Hilbert space $\H$ for which $U-{\mathbb I}$ is compact, originally found
by Kirillov and Ol'shanskii, through constrained quantization of its
coadjoint orbits.  For this purpose the coadjoint orbits are realized
as Marsden-Weinstein quotients.  The unconstrained system, given as a
Weinstein dual pair, is quantized by a corresponding Howe dual pair.
Constrained quantization is then performed in replacing the classical
procedure of symplectic reduction by the $C^*$-algebraic method of
Rief\/fel induction.
Reduction and induction have to be performed with respect to either
$U(M)$, which is straightforward, or $U(M,N)$.  In the latter case one
induces from holomorphic discrete series \rep s, and the desired
result is obtained if one ignores half-forms, and induces from a \rep,
`half' of whose highest weight is shifted relative to the naive orbit
correspondence.  This is only possible when $\H$ is
f\/inite-dimensional.
\end{abstract}

\section{Introduction}
\subsection{Representations from quantized symplectic reduction}
Constrained quantization \ci{Dir,MT} is a very useful method that often
allows one to reduce nonlinear problems in mathematical physics to
linear ones. Such a reduction is possible if a given nonlinear 
(symplectic) space may be written as the reduced (`physical') phase
space relative to a linear phase space with certain constraints def\/ined
on it. The goal of this paper is to quantize the coadjoint orbits
of a certain inf\/inite-dimensional Lie group, which are highly
nonlinear inf\/inite-dimensional symplectic manifolds, by a mathematically
rigorous version of this method. The  group in question (def\/ined
below) has been chosen because it is one of the few
inf\/inite-dimensional Lie groups for which the correspondence between its
irreducible unitary \rep s and its coadjoint orbits is known.
Thus it forms an ideal testing ground for the constrained quantization
(as well as for  more general constructions in mathematical physics)
of inf\/inite-dimensional phase spaces.

 Let $U_0(\H)$ be the Banach Lie group of all unitary operators $U$ on
a separable Hilbert space $\H$ for which $U-{\mathbb I}$ is compact,
equipped with the uniform operator (i.e., norm) topology.  The
continuous unitary \rep s of $U_0(\H)$ were classif\/ied by Kirillov
\ci{Kir1} and Ol'shanskii \ci{Ols1}.  Their classif\/ication
simultaneously applies to the Fr\'{e}chet Lie group $U(\H)$ consisting
of all unitary operators on $\H$, equipped with the strong operator
topology, because all \rep s of $U_0(\H)$ are also strongly
continuous, and can therefore be extended to $U(\H)$. Moreover,
$U(\H)$ re-topologized with the uniform topology has the same
irreducible \rep s on separable Hilbert spaces as the same group
equipped with the strong topology (whose irreducible \rep\ spaces are
automatically separab\-le)~\ci{Pic}. (The \rep\ theory of $U(\infty)$
equipped with the inductive limit topology is much more complicated
\ci{Ols4,Boy2} and will not be discussed here.)

A remarkable aspect of the Kirillov-Ol'shanskii classif\/ication is that
all irreducible \rep s of $U_0(\H)$ may be thought of as the geometric
quantization of certain of its coadjoint orbits.  However, only the
geometric quantization of orbits corresponding to positive eigenvalues
may actually be found in the literature \ci{Boy1}; even this special
case is already fairly involved.  It is this quantization that we
venture to redo, and much simplify, by regarding the orbits as
Marsden-Weinstein quotients, and performing a certain constrained
quantization procedure.

Our work was triggered by Montgomery's observation \ci{Mon} (also cf.\
\ci{LMS}) that for f\/inite-dimensional $\H=\C^k$ certain coadjoint
orbits of $U(k)$ (namely those characterized by positive eigenvalues)
are \MW\ quotients of $\H\ot \C^M$ with respect to $U(M)$, for
suitable $M$ (which depends on the orbit).  The left-action of $U(k)$
and the right-action of $U(M)$ on $\C^k\ot\C^M$ combine to form a
Weinstein dual pair $U(k)\raw \C^k\ot\C^M\leftarrow U(M)$
\ci{KKS,Ste,Wei83}.

 The simplest reduced space thus obtained (viz.\ for $M=1$) is the
projective space $\mathbb P \C^k$; as in the general case, three relevant
symplectic structures, namely its standard form as a K\"{a}hler
manifold, its Lie-Poisson form as a coadjoint orbit, and f\/inally its
Marsden-Weinstein form as a symplectic quotient, all coincide.

We extend Montgomery's result to the situation where the eigenvalues
may be of either sign, and also to the case where $\H$ is
inf\/inite-dimensional. The Weinstein dual pair then~be\-co\-mes
$U_0(\H)\raw \H\ot\C^M\ot\overline{\C}^N\leftarrow U(M,N)$, so that
one reduces with respect to the non-com\-pact group $U(M,N)$. Note that
$M$ and $N$ are f\/inite even in the inf\/inite-dimensional case.

The quantization of the `unconstrained system' $U_0(\H)\raw
\H\ot\C^M\ot\overline{\C}^N\leftarrow U(M,N)$ is trivially done by
Fock space techniques, yielding a Howe dual pair that quantizes the
clas\-sical Weinstein dual pair in question. To quantize the \MW\
reduction pro\-cess that led to the classical coadjoint orbits, we
employ a relatively new method~\ci{NPL93,MT}, which is based on the
\ca ic technique of Rief\/fel induction \ci{Rie74,FD,MT}.  As explained
in \ci{NPL93,MT}, this method in principle quantizes a symplectic
reduction procedure vastly more general than the \MW\ one
\ci{MiW,Xu,NPL93,MT}, and improves on more traditional constrained
quantization techniques (such the Dirac or the BRST method) in cases
where the quantized constraints fail to have a joint eigenvalue
zero. In the context of the present paper, this means that for $N=0$,
where one classically reduces with respect to the compact group
$U(M)$, other techniques would apply as well, whereas for $N>0$ these
would break down.

For $N=0$, the induction procedure is easily carried out on the basis
of Weyl's classical results on tensor products and the symmetric group
\ci{Wey,How4}.  The case $N>0$, where the coadjoint orbit one
quantizes is characterized by eigenvalues of arbitrary sign, is
considerably more complicated.  The quantization of the unconstrained
system $S=\H\ot\C^M\ot\overline{\C}^N$ is known explicitly at least
for f\/inite-dimensional $\H=\C^k$: it is the $k$-fold tensor product of
the metaplectic (or `oscillator', or `Segal-Shale-Weil') \rep\
\ci{Fol}, restricted from $Sp(2(N+M),\R)$ to its subgroup $U(M,N)$
(see \ci{SW,Ste,BR}).

This tensor product has been decomposed by Kashiwara and Vergne
\ci{KV}, also cf.\ Howe~\ci{How2}.  The decomposition of the Hilbert
space quantizing $S=\C^k\ot \C^M\ot\overline{\C}^N$ under $U(k)$ and
$U(M,N)$ does not ref\/lect the decomposition of $S$ under these group
actions if $k>M+N$ (which is the case of relevance to us, as we are
eventually interested in $k=\infty$), cf.\ \ci{Ada2}.  This
fascinating complication implies that for generic coadjoint orbits our
method only works when $\H$ is f\/inite-dimensional.

\subsection{Rief\/fel induction for group actions}
We brief\/ly review how Rief\/fel induction \ci{Rie74,FD,MT} specializes
to the present context. One starts from a strongly Hamiltonian
right-action of a connected Lie group $H$ on a symplectic manifold
$S$, with accompanying equivariant momentum map $J:S\raw \h^*$. We
assume that the reduced space $S^{\mu}\equiv J^{-1}(\O_{\mu})/H$ is a
manifold.

If a Lie group $G$ acts symplectically on $S$ in such a way that its
action commutes with the $H$-action, the reduced space $S^{\mu}$
becomes a symplectic $G$-space in the obvious way; the well-known
`symplectic induction' procedure \ci{KKS,MT} is a special case of this
construction (it is obtained by taking $H\subset G$ and $S=T^*G$).

To quantize the reduced space $S^{\mu}$ and the associated induced
\rep\ of $G$, we assume that a quantization of the unconstrained
system as well as of the constraints are given.  Hence we suppose we
have f\/irstly found a Hilbert space $\F$, which may be thought of as
the (geometric) quantization of $S$.  Secondly, a unitary right-action
(i.e., anti-\rep) $U_R(H)$ on $\F$ should be given, which is the
quantization of the symplectic right-action of $H$ on $S$. Thirdly, we
require a unitary \rep\ $\plc(H)$ on a Hilbert space $\hlc$, which
`quantizes' the coadjoint action of $H$ on the coadjoint orbit
$\O_{\mu}$ This is only possible if the orbit is `quantizable'; for
$H=U(M)$ there is a bijective correspondence between such orbits and
unitary \rep s, and for $U(M,N)$ one obtains at least all unitary
highest weight modules by `quantizing' such orbits \ci{Ada1,Vog}. (In
the latter case the concept of quantization has to be stretched
somewhat to incorporate the derived functor technique to construct
\rep s.)
 
First assuming that $H$ is compact, we construct the induced space
$\huc$ from these data as the subspace of $\F\ot\hlc$ on which
$U_R^{-1}\ot\plc$ acts trivially (here $U_R^{-1}$ is the \rep\ of~$H$
def\/ined by $U_R^{-1}(h)=U_R(h^{-1})$). If $H$ is only locally compact
(and assumed unimodular for simplicity) with Haar measure $dh$, one
has to f\/ind a dense subspace $L\subset \F$ such that the integral
$\int_H dh\,( (U_R^{-1}\ot\plc)(h)\Psi,\Phi)\equiv (\Psi,\Phi)_0$ is
f\/inite for all $\Psi,\Phi\in L\ot\hlc$. This def\/ines a sesquilinear
form $\rip$ on $L\ot\hlc$ which can be shown to be positive
semi-def\/inite under suitable conditions \ci{NPL93}. The induced space
$\huc$ is then def\/ined as the completion of the quotient of $L\ot\hlc$
by the null space of $\rip$; its inner product is, of course, given by
the quotient of $(\cdot ,\cdot )_0$.  For $H$ compact the integral
exists for all $\Psi,\Phi\in\F$ and $(\Psi,\Phi)_0=(P_0\Psi,P_0\Phi)$,
where $P_0$ is the projector onto the subspace of $\F\ot\hlc$ carrying
the trivial \rep\ of $H$, so that we recover the f\/irst description of
$\huc$.

We now assume that a group $G$ acts on $\F$ through a unitary \rep\
 $U_L$; it is required that this action commute with $U_R(H)$.  The
 induced \rep\ $\puc(G)$ on $\huc$ is now def\/ined as follows.  For $H$
 compact, $\puc$ is simply the restriction of $U_L\ot{\mathbb I}$ to
 $\huc\subset \F\ot\hlc$; this is well def\/ined because $U_L\ot{\mathbb
 I}$ commutes with $U_R^{-1}\ot\plc$.  In the general case, one has to
 assume that $U_L$ leaves $L$ stable; then $\puc$ is essentially
 def\/ined as the quotient of the action of $U_L\ot{\mathbb I}$ (on
 $L\ot\hlc$) to $\huc$ as def\/ined above (cf.\ \ci{NPL93} for technical
 details pertinent to the general case).  The Mackey induction
 procedure for group \rep s is recovered by assuming that $H\subset
 G$, and taking $\F=L^2(G)$, cf.\ \ci{Rie74,FD,MT} for details in the
 original setting of Rief\/fel induction, and \ci{NPL93,MT} for the
 above setting.

\section{Representations from Rief\/fel induction}
 In subsections 2.1 to 2.3 we take $\H$ to be an inf\/inite-dimensional
separable Hilbert space, unless explicitly stated otherwise. All
results (sometimes with self-explanatory modif\/ications) are equally
well valid in the f\/inite-dimensional case, which is considerably
easier to handle; we leave this to the reader. We start with the
simplest case, the def\/ining \rep.

\subsection{The quantization of $\PH$}
One can realize $\PH$ as a \MW\ quotient with respect to the group
$U(1)$ \ci{AM,MT}.  Firstly, $\H$ carries a symplectic form $\om$,
expressed in terms of the standard inner product (taken linear in the
f\/irst entry) by $\om(\ps,\phv)=-2\, {\rm Im}\, (\ps,\phv)$.  Secondly,
$U(1)$ (identif\/ied with the unit circle in the complex plane) acts on
$\H$ by $z:\ps\raw z\ps$; this action is symplectic, and yields an
equivariant momentum map \ci{AM} $J:\H\raw {\bf u(1)}^*\equiv \R$
given by $J(\ps)=(\ps,\ps)$.  Then $\PH\simeq J^{-1}(1)/U(1)$.

The quantization of this type of reduced space using Rief\/fel induction
was outlined in the Introduction. We f\/irst need a quantization of the
`unconstrained' system $\H$, which we take to be the symmetric
(bosonic) Fock space $\F=\exp(\H)$ (this is the direct sum of all
symmetrized tensor products $\H^{\ot n}$ ($n=0,1,\ldots$) of $\H$ with
itself). This quantization is so well-established that we will not
motivate it here; cf.\ \ci{Fol,RS1} for mathematical aspects, and
\ci{Woo} for a derivation in geometric quantization.

 The (anti) \rep\ $U_R$ of $U(1)$ on $\F$ is obtained by
`quantization' of the right action on $\H$. We choose $U_R$ as the
second quantization of this right action. Labelling this choice
$U_{R,{\rm sq}}$, this yields $U_{R,{\rm sq}}(z) \res \H^{\ot
n}=z^n{\mathbb I}$. Similarly, the def\/ining \rep\ $U_{1}$ of $G=U(\H)$
(the group of all unitary operators on $\H$) on $\H_1=\H$ yields a
symplectic action on $\H$. This is `second' quantized by the \rep\
$U_{L,{\rm sq}}$ on $\F$, whose restriction $U_n$ to each subspace
$\H^{\ot n}\subset\F$ is the symmetrized $n$-fold tensor product of
$U_{1}$ with itself.  The \rep s $U_{R,{\rm sq}}(U(1))$ and $U_{L,{\rm
sq}}(U(\H))$ obviously commute with each other. Hence $\F$ has a
central decomposition under $U_{L,{\rm sq}}(U(\H))\ot U^{-1}_{R,{\rm
sq}}(U(1))$, which is explicitly given by
\be
\exp(\H)\stackrel{{\rm
sq}}{\simeq} \bigoplus_{n=0}^{\infty} \H_n^{U(\H)}\ot
\overline{\H}_n^{U(1)}.  \ll{dec1}
\ee
Here $\H_n^{U(\H)}$ coincides
with $\H^{\ot n}$, now regarded as the carrier space of the \rep\
$U_n(U(\H))$, which is, in fact, irreducible for all $n$
\ci{Kir1,Ols1} (also cf.\ subsection 3.3 below). Also, ${\H}_n^{U(1)}$
is just $\C$, but regarded as the carrier space of $U_n(U(1))$,
def\/ined by $U_n(z)=z^n$; $\overline{\H}$ stands for the carrier space
of the conjugate \rep.

The general context for decompositions of the type (\ref{dec1}) is the
theory of Howe dual pairs \ci{How1,How3}.  In the present instance,
this applies to $\H=\C^k$, with $U(k)$ and $U(1)$ being the dual pair
in $Sp(2k,\R)$.  (Cf. \ci{Ols4} for the theory of these pairs in the
inf\/inite-dimensional setting.)

 The construction of the induced space $\F^1$ is ef\/fortless in this
 case. The fact that \MW\ reduction took place at $J=1$ means that
 the orbit of $U(1)$ in question is the point $1\in {\bf
 u(1)}^*$. This orbit is quantized by the def\/ining \rep\ $U_1$ of
 $U(1)$ on $\H_1=\C$.  By construction, $\F^1$ is the subspace of
 $\F\ot \H_1=\F$ which is invariant under the \rep\ $U_R^{-1}\ot
 U_1$. Hence (\ref{dec1}) implies that $\F^1=\H$.  The induced \rep\
 $U^1(U(\H))$ on $\F^1$ is simply the restriction of $U_{L,{\rm
 sq}}(U(\H))$ to this space, so that $U^1 \simeq U_{1} $. In other
 words, we have recovered the def\/ining \rep.

So far, so good, but unfortunately there is a subtlety if one derives
$U_R$ and $U_L$ from geometric quantization. Using the `uncorrected'
formalism (as described, e.g., in Ch.\ 9 of~\ci{Woo}), exploiting the
existence of an invariant positive totally complex polarization, viz.\
the anti-holomorphic one, one f\/inds that $\F$ is realized as the space
of holomorphic functions on $\H$. The quantization $\pq$ of the
momentum maps $J_R$ for $U(1)$ and $J_L$ for $U(\H)$ (with respect to
their respective actions on $\C^k$) then reproduces the second
quantizations $U_{R,{\rm sq}}$ and $U_{L,{\rm sq}}$, respectively.

 If, however, one is too sophisticated and incorporates the half-form
correction to geometric quantization \ci[Ch.\ 10]{Woo}, one obtains
extra contributions: for $\H=\C^k$, $\pq(J_R)$ is replaced by
$\pq(J_R)+k/2$, whereas $\pq(J_L)$ acquires an additional constant
$\half$ (times the unit matrix). These Lie algebra \rep s exponentiate
to unitary \rep s of double covers $\til{U}(k)$ and $\til{U}(1)$,
which we denote by $U_{L,{\rm hf}}$ and $U_{R,{\rm hf}}$,
respectively. Under $U_{L,{\rm hf}}(\til{U}(k))\ot U^{-1}_{R,{\rm
hf}}(\til{U}(1))$ we then f\/ind the central decomposition
\be
\exp(\H)\stackrel{{\rm hf}}{\simeq} \bigoplus_{n=0}^{\infty}
\H_{(n+\half,\half,\ldots,\half)}^{\til{U}(k)}\ot
\overline{\H}_{n+\half k}^{\til{U}(1)}. \ll{dec2}
\ee
Here
$\H_{(n+\half,\half,\ldots,\half)}$ carries the \rep\ of $\til{U}(k)$
with highest weight $ (n+\half,\half,\ldots,\half)$; this is the
tensor product of $\H_n$ and the square-root of the determinant
\rep. One observes that the inclusion of half-forms is awkward for
Rief\/fel induction -- we defer a discussion of this point to Chapter 3.

\subsection{The coadjoint orbits of $U_0(\H)$ as reduced spaces}
The Lie algebra $\g={\bf u}_0(\H)=i{\mathfrak K}(\H)_{\rm sa}$ of
$G=U_0(\H)$ consists of all skew-adjoint compact operators on $\H$
with the norm topology. The dual $\g^*={\bf u}_0(\H)^*$ is the space
of all self-adjoint trace-class operators on $\H$, with topology
induced by the trace norm $\n \rh\n_1={\rm Tr}\, |\rh|$ (this
coincides with the weak$\mbox{}^*$ topology). The pairing is given by
$\la\rh,X\ra=i\, {\rm Tr}\, \rh X$.

The coadjoint action of $U_0(\H)$ on $\lu$ is given by
$\pco(U)\rh=U\rh U^*$. We are interested in those coadjoint orbits
which are `quantizable' in the sense of geometric quantization, since
their quantization should produce all irreducible \rep s of $U_0(\H)$
\ci{Kir1,Kir2}.  Each such orbit is labeled by a pair $({\sf m}, {\sf
n})$, where ${\sf m}$ is an ordered $M$-tuple of positive integers
satisfying $m_1\geq m_2\geq\ldots m_M>0$, and $\sf n$ is a similar
$N$-tuple ($M,N<\infty$). The coadjoint orbit $\Omn$ consists of all
elements of $\lu$ with eigenvalues
$m_1,m_2,\ldots,m_M,0^{\infty},-n_N,\ldots,-n_1$. The degeneracy of
each numerical eigenvalue $m_i$ (or $-n_j$) is simply the number of
times it occurs in this list. The explicit quantization of the orbits
$\Omn$ is not discussed in \ci{Kir1,Kir2}; the case where either $\sf
m$ or $\sf n$ is empty is done in \ci{Boy1} using geometric
quantization.

For f\/inite-dimensional $\H$, it was shown by Montgomery \ci{Mon} that
$\O_{{\sf m},0}$ can be written as a \MW\ reduced space with respect
to the natural right-action of $U(M)$ on $\H\ot\C^M$.  This is a
special instance of the theory of dual pairs.  With $\H=\C^k$, the
groups $U(\H)$ and $U(M)$ form a Howe dual pair inside the symplectic
group $Sp(2kM,\R)$ \ci{How1,Ste,How3}, and the momentum maps $J_R$ and
$J_L$ introduced below build a Weinstein dual pair, cf.\
\ci{KKS,Wei83}. General theorems on the connection between coadjoint
orbits of one group and \MW\ reduced spaces with respect to the other
group in a dual pair are given in \ci{LMS}. We will now generalize the
special case mentioned above to inf\/inite-dimensional $\H$, and general
orbits $\Omn$.

We take $S=\H\ot\C^{M+N}$, which we regard as a Hilbert manifold in
the obvious way.  We choose the canonical basis
$\{e_i\}_{i=1,\ldots,M+N}$ in $\C^{M+N}$. The symplectic form $\om$ on
$S$ is taken as (we put $\pl =1$)
\be
\om(\ps,\phv)=-2\, {\rm Im}\,\left
( \sum_{i=1}^M(\ps_i,\phv_i)-\sum_{i=M+1}^{M+N}(\ps_i,\phv_i)\right) ,
\ll{ommn}
\ee
where we have expanded $\ps=\sum_i\ps_i\ot e_i$ and
similarly for $\phv$.  It is convenient to introduce an indef\/inite
sesquilinear form on $\C^{M+N}$ by putting $(e_i,e_j)=\pm \dl_{ij}$,
with a plus sign for $i=1,\ldots,M$ and a minus sign for
$i=M+1,\ldots, M+N$. Together with the inner product on $\H$ this
induces an indef\/inite form $(\cdot,\cdot)_S$ on $S$ in the obvious
(tensor product) way. The right-hand side of (\ref{ommn}) then simply
reads $-2\, {\rm Im}\, (\ps,\phv)_S$. A simple trick shows that $S$ is
strongly symplectic: we can regard $S$ as a Hilbert space
$\H\ot\C^M\oplus \overline{\H\ot\C^N}$, with inner product
$(\ps,\phv)_{\mbox{\footnotesize\rm trick}}=\sum_{i=1}^M(\ps_i,\phv_i)+
\sum_{i=M+1}^{M+N}(\phv_i,\ps_i)$.  Then $\om(\ps,\phv)=-2\, {\rm
Im}\,(\ps,\phv)_{\mbox{\footnotesize\rm trick}}$, and the claim follows from
the well-known fact that Hilbert spaces are strongly symplectic~\ci{AM}.

 The Lie group $H=U(M,N)$ (which is $U(M)$ or $U(N)$ for $\sf n$ or
$\sf m$ empty) acts on $S$ from the right in the obvious way, i.e., by
$U\raw\I\ot U^T$. This action is symplectic, with anti-equivariant
momentum map $J_R:S\raw (\h^*)^-$.  If we identify $X\in \h$ with a
generator in the def\/ining \rep\ of $H$ on $\C^{M+N}$, we obtain (cf.\
\ci[p.\ 501]{KKS})
\be
\la J_R(\ps),X\ra = i(\I\ot X^T\ps,\ps)_S.
\ll{jr}
\ee
On a suitable Cartan subalgebra $\mathfrak t$ of $\h$, which
we identify as the set of imaginary diagonal operators on $\C^{M+N}$,
with basis $H_j=-iE_{jj}$, this simply reads $\la J_R(\ps),H_j\ra =
\pm (\ps_j,\ps_j)$ with a plus sign for $j=1,\ldots,M$ and a minus
sign for $j=M+1,\ldots, M+N$.

We now identify $\wmn$ with an element of $\h^*$ by the pairing $\la
\wmn,X\ra=i{\rm Tr}\, D_{\wmn}X$, where $ D_{\wmn}$ is the diagonal
matrix in $M_{M+N}(\C)$ with entries $m_1,\ldots, m_M,
-n_N,\ldots,-n_1$. This means that $\wmn$ def\/ines a dominant integral
weight on $\mathfrak t$, and vanishes on its complement.  The subset
$J_R^{-1}(\wmn)$ of $S$ consists of those vectors $\ps=\sum_i\ps_i\ot
e_i$ for which $(\ps_i,\ps_i)=m_i$ for $i=1,\ldots, m$, and
$(\ps_{M+j},\ps_{M+j})=n_{N+1-j}$ for $j=1,\ldots, n$, with the
$\ps_k$'s mutually orthogonal. The normalizations come from $J_R$
evaluated on $\mathfrak t$, and the orhtogonality derives from the
constraint that $J_R$ vanish on its complement.  {\em Note that the
integrality of the $m_i$ and $n_j$ plays no role in this subsection.}

\begin{lemma}
$J_R^{-1}(\wmn)$ is a submanifold of $S$. \ll{subm}
\end{lemma}
{\em Proof.}  According to the theorem on p.\ 550 of \ci{AMP}, we need
to show that $J_R: J_R^{-1}(\wmn)\raw \h^*$ is a submersion, which is
the case if at any point $\ps\in J_R^{-1}(\wmn)\subset S$ the
derivative $(J_R)_*\equiv J_R^{(1)}:T_{\ps}S\raw
T_{J_R(\ps)}\h^*\simeq \h^*$ is surjective and has a complementable
kernel.  The former is equivalent to the statement that $\ps$ is a
regular value of the momentum map \ci{AM}.  The derivative at $\ps\in
S$ follows from (\ref{jr}) as
\be \la (J_R^{(1)})_{\ps}(\xi),X\ra= 2
{\rm Re}\, (\I\ot iX^T\xi,\ps)_S. \ll{derjr}
\ee
This formula shows
that $J_R^{(1)}$ is continuous, so that its kernel is closed. The
complementability of this kernel is then immediate, since $S$ is a
Hilbert manifold.  The surjectivity of $J_R^{(1)}$ follows from
(\ref{derjr}) by inspection, but it is more instructive to derive it
from Prop.\ 2.11 (due to Smale) in \ci{Mar}. This states that $\ps$ is
a regular value of the momentum map if\/f the stability group
$H_{\ps}\subseteq H$ of $\ps$ is discrete.  Now, as pointed out
earlier, $\ps=\sum_i \ps_i\ot e_i \in J_R^{-1}(\wmn)$ implies that all
$\ps_i$ are nonzero are orthogonal, so that $H_{\ps}$ is just the
identity.  \hfill $\blacksquare$

\medskip

The action of $H$ on $S$ is not proper unless $\sf m$ or $\sf n$ is
empty (in which case $H$ is compact). However:

\begin{lemma} The action of $H$ on $J_R^{-1}(\wmn)$ is proper.
\end{lemma}
{\em Proof.} Let $\ps^{(n)}\raw \ps$ in $S$; equivalently,
$\ps_i^{(n)}\raw \ps_i$ in $\H$ for all $i$. If $\{U^{(n)}\}$ is a
sequence in $H$ and $U^{(n)}\ps^{(n)}$ converges, the fact that for
each $n$ all $\ps_i^{(n)}$ are nonzero and orthogonal implies that
$\{U_{ij}^{(n)}e_j\}$ must converge in $\C^{M+N}$ for each $i$. Since
convergence in the topology on $U(M,N)$ is given by convergence of all
matrix elements in the def\/ining representation, this implies that
$\{U^{(n)}\}$ must converge in $H$. \hfill $\blacksquare$

\medskip

By the standard theory of \MW\ reduction \ci{Mar74,AM}, these lemmas
imply that the reduced space
\be
S^{\wmn}=J_R^{-1}(\wmn)/H_{\wmn}
\ll{sred}
\ee
(where $H_{\wmn}$ is the stability group of
$\wmn\in\h^*$ under the coadjoint action) is a smooth symplectic
manifold. We will proceed to show that it is symplectomorphic to the
coadjoint orbit $\Omn\in\g^*$, where $G=U_0(\H)$, as explained above.
The required dif\/feomorphism is given by a quotient of the momentum map
$J_L:S\raw \g^*$ def\/ined from the natural left-action of $G$ on $S$,
which action is evidently symplectic. Identifying $\g$ with the space
of compact skew-adjoint operators $Y$ on $\H$, one easily f\/inds that
this momentum map is given by
\be -i\la J_L(\ps),Y\ra = (Y\ot {\mathbb I}
\ps,\ps)_S
=\sum_{i=1}^M(Y\ps_i,\ps_i)-\sum_{i=M+1}^{M+N}(Y\ps_i,\ps_i).  \ll{jl}
\ee
Since the left-$G$ action and the right-$H$ action commute, $J_L$
is invariant under $H$ (i.e., $J_L(\ps U)=J_L(\ps)$ for all $U\in H$
and $\ps\in \H$), so that $J_L$ (restricted to $J_R^{-1}(\wmn)$)
quotients to a well-def\/ined map $\jlt:S^{\wmn}\raw\Omn$. Once we have
shown that $\jlt$ is a dif\/feomorphism, it will follow that it is
symplectic, because of the def\/inition of the symplectic structure on
$S^{\wmn}$ and the fact that $J_L$ is equivariant.

 Generalizing a standard result in the root and weight theory for
compact Lie groups, see e.g.\ \ci{Kna}, we f\/irst note that the the
stability group of $\wmn\in \h^*$ under the coadjoint action is
$H_{\wmn}=\prod_l U(l)$, where $\sum l=M+N$, and the product is over
the multiplicities within either $\sf m$ or $\sf n$ in $\wmn$; this is
a subgroup of $U(M,N)$ in the obvious block-diagonal form.  (For
example, if $\wmn =((2,1,1),(2,2,2))$ the stability group is
$U(1)\times U(2)\times U(3)$.)  It then follows from (\ref{jl}) that
$\jlt$ is a bijection onto $\Omn$.

\begin{prop}
$\jlt$ is smooth. \ll{s1}
\end{prop}
{\em Proof.} The manifold structure of $\Omn$ is def\/ined by its
embedding in $\g^*$, which is a Banach space in the trace-norm
topology (cf.\ the beginning of this section).  The smoothness of
$\jlt$ then follows from that of $J_L:J_R^{-1}(\wmn)\raw\g^*$, since
the Lie group $H$ acts smoothly, freely, and properly on
$J_R^{-1}(\wmn)$.

{\em 1. Continuity of $J_L$.}  We prove continuity on all of $S$. As a
map between separable metric spaces ($S$ is separable because $\H$ is
by assumption, and $\g^*$ is separable because the f\/inite-rank
operators are dense in it), $J_L$ is continuous if $\ps^{(n)}\raw \ps$
in $S$ implies $J_L(\ps^{(n)})\raw J_L(\ps)$ in $\g^*$. The topology
on $\g^*$ coincides with the weak$\mbox{}^*$-topology, so the desired
continuity follows from (\ref{jl}), the boundedness of $Y$, and
Cauchy-Schwartz.

{\em 2.  Existence and continuity of $J_L^{(1)}$.} The derivative of
 $J_L$ at $\ps$ is given by
 \be \la (J_L^{(1)})_{\ps}(\xi),Y\ra= 2
 {\rm Re}\, \left(
 \sum_{i=1}^M(iY\xi_i,\ps_i)-\sum_{i=M+1}^{M+N}(iY\xi_i,\ps_i)\right).
 \ll{derjl}
 \ee
 By the same reasoning as in the previous item,
 $(J_L^{(1)})_{\ps}$ lies in ${\cal L}(S,\g^*)$ and is continuous.

The second derivative $J_L^{(2)}:S\times S\raw \g^*$ can be read of\/f
from (\ref{derjl}); its existence and continuity are established as
before. Higher derivatives vanish. \enp

\begin{prop}
$\jlt^{-1}$ is smooth. \ll{s2}
\end{prop}
{\em Proof.} We pick an arbitrary point $\rh_0\in\Omn$, with stability
group $G_0$. Let $\H=\oplus_l\H_l$ be the decomposition of $\H$ under
which $\rh_0$ is diagonal (the dimension of each $\H_0$ is the
degeneracy of the corresponding eigenvalue; this dimension is f\/inite
unless the eigenvalue is 0). Then $G_0=\oplus_l U_0(\H_l)$, in
self-evident notation. The Lie algebra $\g_0$ of $G_0$ is given by
those operators in $\g= i{\mathfrak K}(\H)_{\rm sa}$ which commute with
$\rh_0$. The manifold $\Omn$ is modelled on $\g/\g_0$. This has the
quotient topology inherited from $\g$, i.e., the trace-norm topology
determined by $\n A\n_1={\rm Tr}\, |A|$.

We def\/ine a neighbourhood $V_0\subset \Omn$ of $\rh_0$ as
follows. Since $G$ is a Banach-Lie group, by \ci{LT} there exists a
neighbourhoud $V$ of $0\in\g$ such that $\exp$ is a dif\/feomorphism on
$V$ into $\g$.  We put $V_0=\{\pco(\exp(A))\rh_0| A\in V\}$ (recall
that the coadjoint action is given by $\pco(U)\rh=U\rh U^*$). To
def\/ine a chart on $V_0$, we f\/irst show that $\g$ (equipped with the
trace-norm topology) admits a splitting $\g=\g_0\oplus \m_0$.  Here
$\m_0$ consists of those operators~$A$ in $\g$ whose matrix elements
$(A\ps,\phv)$ vanish if both $\ps$ and $\phv$ lie in the same space
$\H_l$, for all $l$. It is clear that $\g=\g_0\oplus \m_0$ as a set,
and it quickly folows that each summand is closed: since $\n A\n\leq
\n A\n_1$, the uniform topology is weaker than the trace-norm one, so
that closedness in the former implies the corresponding property in
the latter topology.  As to the uniform closedness of $g_0$, one has
$\n [A,\rh_0]\n\,\leq 2 \n A\n\; \n \rh_0\n$, so that $\g_0\ni A_n\raw
A$ implies that $A\in \g_0$. On $\m_0$ an even more elementary
inequality does the job. Thus $\g/\g_0\simeq \m_0$, and we may use
$\m_0$ as a modelling space for $\Omn$.

We def\/ine a chart on $V_0$ by $\phv_0:V_0\raw \m_0$, given by
$\phv_0(\pco(\exp(A))\rh_0)=A_0$, where $A_0$ is the component of
$A\in\g$ in $\m_0$.  We would like to model $S^{\wmn}$ on $\m_0$ as
well, but this is not directly possible because it has the wrong
topology. Hence the following detour.  Take a $\ps_0\in
J_R^{-1}(\wmn)\subset S$ for which $J_L(\ps_0)=\rh_0$.  Using the fact
that $J_L$ is a bijection, we model $S^{\wmn}=J_R^{-1}(\wmn)/H_{\wmn}$
on the closed linear subspace of $S$ given by $M_0=\{A\ot\I \ps_0|A\in
\m_0\}$, equipped with the relative topology of $S$.  Put
$W_0=\{\exp(A)\ot\I \ps_0|A\in m_0\}\subset S$. If $pr:J_R^{-1}\raw
J_R^{-1}(\wmn)/H_{\wmn}$ is the canonical projection, we have a chart
on the neighbourhood $pr(W_0)$ of $pr(\ps_0)$ def\/ined by
$\ph_0:pr(W_0)\raw M_0$ given by $\ph_0(pr(\exp(A)\ps_0))=A\ps_0$.
This procedure respects the manifold structure of $S^{\wmn}$, which by
def\/inition is quotiented from $J_R^{-1}(\wmn)\subset S$.

We now def\/ine
$\mbox{}_0\jlt^{-1}=\ph_0\circ\jlt^{-1}\circ\phv_0^{-1}$; this is a
map from $\phv_0 (V_0)\subset \m_0$ to $\ph_0\circ pr(W_0)\subset
M_0$. Clearly, $\mbox{}_0\jlt^{-1}(A)=A\ps_0$.  This immediately
implies that $\mbox{}_0\jlt^{-1}$, and therefore $\jlt^{-1}$, is
smooth.  \enp

To sum up, we have proved
\begin{theorem}
For any separable Hilbert space $\H$, the coadjoint orbit $\Omn$ of
the group $U_0(\H)$ (which consists of all trace-class operators on
$\H$ with $M$ specif\/ic positive and $N$ specific negative eigenvalues)
is symplectomorphic to the \MW\ quotient
$S^{\wmn}=J_R^{-1}(\wmn)/H_{\wmn}$ with respect to $S=\H\ot\C^{M+N}$
and the natural right-action of $H=U(M,N)$. \ll{omw}
\end{theorem}

\subsection{Representations induced from $U(M)$}
The \rep s of $U_0(\H)$ were fully classif\/ied in \ci{Kir1,Ols1,Ols3}
(also cf.\ \ci{Kir2,Ols4,Boy2}).  A~re\-mar\-kable fact is that $U_0(\H)$
is a type I group, so that all its factorial \rep s are of the form
$U\ot\I$ on $\H_{U}\ot\H_{\rm mult}$, where $(U,\H_{U})$ is
irreducible.  Each \irep\ cor\-responds to an integral weight $\wmn$ of
the type specif\/ied above, where $M$ and $N$ are arbitrary (but
f\/inite). The carrier space $\Hmn$ is of the form $\Hmn=\Hm\ot \Hn$,
and car\-ries the \irep\ $\pmn=\pim\ot\pin$. Here $\Hm$ is the subspace
of $\ot^M\H$ obtained by symmetrization according to the Young diagram
whose $k$-th row has length $m_k$, and $\Hn$ is the conjugate space of
$\H^{{\sf n}}$. The \rep\ $\pim$ is the one given by the restriction
of the $M$-fold tensor product of the def\/ining \rep\ of $U_0(\H)$ to
$\Hm$,~etc.

 This is almost identical to the theory for f\/inite-dimensional
$\H=\C^k$ \ci{Wey,Zel} (which has the obvious restriction that
$M,N\leq k$); the only dif\/ference is that in the inf\/inite-dimensional
case $\Hm\ot\Hn$ is already irreducible. For $k<\infty$, on the other
hand, one needs to take the so-called Young product \ci{Zel} of $\Hm$
and $\Hn$ rather than the tensor product (this is the irreducible
subspace generated by the tensor product of the highest-weight vectors
in each factor); moreover, the use of conjugate spaces may be avoided
in that case by tensoring with powers of the determinant \rep. For
example, $\C^k\ot \overline{\C}^k$ contains the irreducible subspace
$\sum_{i=1}^k e_i\ot \overline{e_i}$ which does not lie in the Young
product; for $k=\infty$ this subspace evidently no longer exists.  For
$M=0$ or $N=0$ there is no dif\/ference whatsoever.

We will now show how the \rep s $(\pim,\Hm)$ can be obtained by
Rief\/fel induction; the \rep s $(\pin,\Hn)$ may then be constructed
similarly. This will quantize the coadjoint orbits $\O_{\sf m}\equiv
\O_{({\sf m},\emptyset)} $ and $\O^-_{\sf n}\equiv\O_{(\emptyset,{\sf
n})}$, respectively.  We note that $\O^-_{\sf n}$ is $\O_{\sf n}$ with
the sign of the symplectic form changed; this relative minus sign
corresponds to the passage from $\H$ to $\overline{\H}$ upon
quantization.

Our starting point is Theorem \ref{omw}, in which we take
$S=\H\ot\C^M$, with $H=U(M)$ acting on $S$ from the right and
$G=U_0(\H)$ acting from the left in the natural way; we call these
actions $U_1^T(H)$ and $U_1(G)$, respectively. As explained in the
Introduction, we f\/irst have to quantize $S$ and the group actions
def\/ined on it. We do so by taking the bosonic second quantization, or
symmetric Fock space, $\F=\exp(S)$ over $S$ \ci{RS1,Woo}, cf.\
subsection~2.1.  For later use, we equivalently def\/ine this as the
subspace of $\sum_{n=0}^{\infty} \ot^n S$ on which the natural \rep\
of the symmetric group $S_n$ on $\ot^n S$ acts trivially for all $n$.

As in the $M=1$ case (cf.\ subsection 2.1) we f\/irst investigate the
\rep s of $U_0(\H)$ and $U(k)$ on $\F$ obtained by second
quantization, or equivalenty, by geometric quantization without the
half-form modif\/ication. This goes as follows.  The groups $H$ and $G$
act on each subspace $\ot^n S$ by the $n$-fold tensor product of their
respective actions on $S$, and these actions restrict to $\F$.  Thus
the actions $U_1^T(H)$ (which we turn into a \rep\ by taking the
inverse) and $U_1(G)$ on $S$ are quantized by the unitary \rep s
$\Gamma\overline{U}_1(H)$ ($=U_{R,{\rm sq}}^{-1}(H)$ in the notation
of subsection 2.1, and $U_R^{-1}(H)$ in that of the Introduction) and
$\Gamma U_1(G)$ ($=U_{L, {\rm sq}}(G)$), respectively (note that
$U^T_1(h^{-1})=\overline{U}_1(h)$).  Here $\Gamma$ is the second
quantization functor \ci{RS1}. This setup, and the associated central
decomposition of $\F$ under these group actions, illustrate Howe's
theory of dual pairs \ci{How1,How2,How3} in an inf\/inite-dimensional
setting, cf.\ \ci{Ols4}.

The fact that the coadjoint orbit $\O_{\sf m}$ of $G$ is
(symplectomorphic to) the \MW\ quotient of $S$ with respect to ${\sf
m}\in\h^*$, cf.\ Theorem \ref{omw}, should now be ref\/lected, or rather
quantized, by constructing the unitary \rep\ $\pim(G)$ (which
according to Kirillov is attached to $\O_{\sf m}$) by Rief\/fel
induction from the \rep\ $\plm(H)$ attached to the orbit through $\sf
m$ in $H$.  Here $\plm(U(M))$ is simply the unitary \irep\ given by
the highest weight $\sf m$; it is realized on $\Hlm$, which is the
subspace of $\ot^M \C^M$ obtained by symmetrization according to the
Young diagram whose $k$-th row has length~$m_k$.

To f\/ind the carrier space of the induced \rep\ $\pim(G)$ we merely
have to identify the subspace of $\F\ot\Hlm$ which is invariant under
$\Gamma\overline{U}_1\ot\plm(H)$. This is very easy on the basis of
the following well-known facts \ci{Wey,Zel,How4}:
\begin{enumerate}
\item
%1
The \rep s of the symmetric group $S_n$ are self-conjugate; for any
\irep\ $\pll(S_n)$, the tensor product $\pll\ot\pll$ contains the
identity \rep\ once, and $\pll\ot \plla$ does not contain the identity
unless ${\sf l}={\sf l}'$. (Recall that the \irep s of $S_n$ are
labelled by an $n$-tuple of integers ${\sf l}=(l_1,\ldots,l_n)$, where
$l_1\geq l_2\geq \ldots l_n\geq 0$ and $\sum_i l_i=n$.)  The
collection of all such $n$-tuples $\sf l$ forms the dual $\hat{S}_n$.
\item
%2
Any unitary \irep\ $\pll(U(M))$ is given by an $M$-tuple ${\sf
l}=(l_1,\ldots,$ $l_M)$ of positive nondecreasing integers (possibly
zero), as in the preceding item, or by the conjugate $\overline{\pll}$
of such a \rep. Then $\pll\ot\overline{\pll}$ contains the identity
\rep\ once, but the identity does not occur in any $\pll\ot\plla$, or
in any $\pll\ot\overline{\plla}$ unless in the latter case ${\sf
l}={\sf l}'$.
\item
%3
The def\/ining \rep\ of $S_n$ on $\ot^n \C^M$ commutes with the $n$-fold
tensor product of the conjugate of the def\/ining \rep\ of $U(M)$, so
that one has the central decomposition
\be
\ot^n \C^M \simeq \bigoplus_{{\sf l}'\in \hat{S}_n}
\Hll^{S^n}\ot \overline{\H}_{\sf l}^{U(M)}, \ll{cd1}
\ee
where the prime (relevant only when $M<n$) on
the $\oplus$ indicates that the sum is only over those $n$-tuples $\sf
l$ for which $l_{M+1}=0$. Here $\Hll^{S^n}$ is the carrier space of
$\pll(S^n)$, and $\overline{\H}_{\sf l}^{U(M)}$ is the carrier space
of the conjugate of the \irep\ of $U(M)$ obtained by making $\sf l$ an
$M$-tuple by adding or removing zeros. (A simliar statement holds
without the conjugation, of course.)
\item
%4
Similarly,
\be \ot^n \H \simeq \bigoplus_{{\sf l}\in \hat{S}_n}
\Hll^{S^n}\ot \Hm, \ll{cd2}
\ee
under the appropriate \rep s of $S_n$
and $U_0(\H)$, where $\Hm$ was introduced at the beginning of this
subsection (for $\H=\C^k$ this is equivalent to a classical result in
invariant theory, see e.g.  \ci[4.3.3.9]{How4}).
\end{enumerate}

 Now consider $\ot^n (\H\ot\C^M)\simeq \ot^n \H\,\ot\, \ot^n
\C^M$. This carries the \rep\ $U_n^{\H}\ot U_n^{\C^M}$ of $S_n$, where
$U_n^{\cal K}(S_n)$ is the natural \rep\ on $\ot^n {\cal K}$. Applying
items 4 and 3, and subsequently 1 above, we f\/ind that the subspace
$\ot^n_s(\H\ot\C^M)\subset \ot^n (\H\ot\C^M)$ which is invariant under
$S_n$ can be decomposed as
\be
\bigotimes^n_s(\H\ot\C^M)\simeq
\bigoplus_{{\sf l}'\in \hat{S}_n} \H^{\sf l}\ot \overline{\H}_{\sf
l}^{U(M)}, \ll{decohil}
\ee
in the sense that the restriction $\ot^n_s
(U_1(G)\ot\overline{U}_1(H))$ of $\Gamma U_1(G)\ot
\Gamma\overline{U}_1(H)$ (def\/ined on $\F=\exp(\H\ot\C^M)$) to
$\ot^n_s(\H\ot\C^M)\subset \F$ decomposes as
\be
\bigotimes^n_s (U_1(G)\ot\overline{U}_1(H))\simeq
\bigoplus_{{\sf l}'\in \hat{S}_n}
U^{\sf l}(G)\ot \overline{\pll}(H). \ll{decorep}
\ee
We then apply
item 2 to conclude that the only subspace of $\F\ot\Hlm$ which is
invariant under $\Gamma\overline{U}_1\ot\plm(H)$ corresponds to
$n=\sum_{i=1}^M m_i$ (where $m_i$ are the entries of the $M$-tuple
$\sf m$). Moreover, by (\ref{decohil}) this invariant subspace is
exactly $\Hm$ as a $U_0(\H)$ module. Hence we have proved
\begin{theorem}
Regard the symmetric Fock space $\F=\exp(\H\ot\C^M)$ as a left-module
(\rep\ space) of $U_0(\H)$ and a right-module of $U(M)$ under the
second quantization of their respective natural actions on
$\H\ot\C^M$. Applying Rieffel induction to this bimodule, inducing
from the \irep\ $\plm(U(M))$ (which corresponds to the highest weight
${\sf m}=(m_1,\ldots,m_M)$), yields the induced space $\Hm$ carrying
the \irep\ $\pim(U_0(\H))$. \ll{main}
\end{theorem}
This, then, is the exact quantum counterpart of Theorem \ref{omw},
specialized to ${\sf n}=\emptyset$.  As remarked earlier, there exists
an obvious analogue of Theorem \ref{main} for ${\sf m}=\emptyset$, in
which all Hilbert spaces and \rep s occurring in the construction are
replaced by their conjugates.

  To prepare for the next subsection we will now give a slight
reformulation of the proof.  We start with f\/inite-dimensional
$\H=\C^k$, with $k>M$. Classical invariant theory \ci{How4} then
provides the decomposition of $\exp(S)$ under $\Gamma U_1(U(k))\ot
\Gamma \overline{U}_1(U(M))$ as
\be
\exp(\C^k\ot\C^M) \stackrel{{\rm
sq}}{\simeq} \bigoplus_{{\sf l}\in D_M} \H_{\sf l}^{U(k)}\ot
\overline{\H}_{\sf l}^{U(M)}, \ll{howe}
\ee
where the sum is over all
Young diagrams (or tuples) $D_M$ with $M$ rows or less, including the
empty diagram. (Note that it would have been consistent with our
previous notation to write $(\H^{\sf l})^{U(k)}$ for $\H_{\sf
l}^{U(k)}$; both stand for the irreducible \rep\ of $U(k)$ def\/ined by
the Young diagram $\sf l$.  In what follows, we will reserve the
notation $\Hul$ for $\Hll(U_0(\H))$, where $\H=l^2$.) Eq.\
(\ref{howe}) is an illustration of the theory of Howe dual pairs
\ci{How1,How2,How3}: it exhibits a multiplicity-free central
decomposition of $\F=\exp(S)$ under the commuting actions of $U(k)$
and $U(M)$ (which form a dual pair in $Sp(2kM,\R)$, of which $\F$
carries the metaplectic \rep).

In order to study the limit $k\raw\infty$ we realize $\exp(\H\ot\C^M)$
(with $\H=l^2$ now inf\/inite-dimesional) as an (incomplete) inf\/inite
tensor product \ci{vonN} with respect to the vacuum vector
$\Omega\in\exp(\C^M)$, that is (recalling $\exp(\C^k\ot\C^M)\simeq
\ot^k \exp(\C^M)$), $\exp(\H\ot\C^M)\simeq
\ot_{\Omega}^{\infty}\exp(\C^M)$, where the right-hand side is the
Hilbert space closure (with respect to the natural inner product on
tensor products) of the linear span of all vectors of the type
$\ps_1\ot\ldots \ps_l\ot\Omega\ot\Omega\ldots$, $\ps_i\in\exp(\C^M)$,
in which only f\/initely many entries dif\/fer from $\Omega$. (The term
`incomplete' refers to the fact that only `tails' close to an inf\/inite
product of $\Om$'s appear.)  Thus $\exp(\C^k\ot\C^M)\simeq \ot^k
\exp(\C^M)$ is naturally embedded in $\exp(\H\ot\C^M)$ by simply
adding an inf\/inite tail of $\Om$'s, and this provides an embedding
$\exp(\C^k\ot\C^M)\subset \exp(\C^{k+1}\ot\C^M)$ as well. Clearly,
$\exp(\H\ot\C^M)$ coincides with the closure of the inductive limit
$\cup_{k=1}^{\infty} \exp(\C^k\ot\C^M)$ def\/ined by this embedding.

 Choosing the natural basis in $\H=l^2$, we obtain an embedding
$U(k)\subset U(k+1)$, with corresponding actions on $\H$; our group
$U_0(\H)$ (realized in its def\/ining \rep\ on $\H$) is the norm-closure
of the inductive limit group $\cup_{k=1}^{\infty} U(k)$. Using the
explicit realization of $\H^{\sf l}$ as a Young-symmetrized tensor
product, we similarly obtain embeddings $\Hll(U(k))\subset
\Hll(U(k+1))$. Thus the inductive limit $\cup_{k=1}^{\infty}
\Hll(U(k))$ is well-def\/ined. Using~(\ref{howe}), we then have that
$\exp(\H\ot\C^M)$ is the closure of $\cup_{k=1}^{\infty} \oplus_{{\sf
l}\in D_M} \H_{\sf l}^{U(k)}\ot \overline{\H}_{\sf l}^{U(M)}$, which
in turn coincides with the closure of $\oplus_{{\sf l}\in
D_M}\cup_{k=1}^{\infty}\H_{\sf l}^{U(k)}\ot \overline{\H}_{\sf
l}^{U(M)}$. We now use the fact that the closure of
$\cup_{k=1}^{\infty}\H_{\sf l}^{U(k)}$ is $\Hul$ as a \rep\ space of
$U_0(\H)$ (this is obvious given the explicit realization of these
spaces, but it is a deep result that an analogous fact holds for all
\rep s of $U_0(\H)$ \ci{Ols1,Ols3,Ols4}).  This yields the desired
decomposition
\be
\exp(\H\ot\C^M) \stackrel{{\rm sq}}{\simeq}
\bigoplus_{{\sf l}\in D_M} \H_{\sf l} \ot \overline{\H}_{\sf
l}^{U(M)}, \ll{howeinf}
\ee
under $\Gamma U_1(U_0(\H))\ot \Gamma
\overline{U}_1(U(M))$. This result was previously derived in \ci{Ols4}
using a technique of holomorphic extension of \rep s.

Starting from (\ref{howeinf}), Theorem \ref{main} follows immediately
from item 2 on the list of ingredients of our previous proof.

To end this subsection we register how the half-form correction to
geometric quantization modif\/ies (\ref{howe}), cf.\ subsection 2.1, and
in particular (\ref{dec2}). These corrections are f\/inite only for
$\H=\C^k$, $k<\infty$, so we only discuss that case.  As for $M=1$,
one f\/inds that the half-form quantizations of the momentum maps
corresponding to the $U(k)$ and $U(M)$ actions on $\C^k\ot \C^M$ lead
to Lie algebra \rep s that can only be exponentiated to \rep s
$U_{L,{\rm hf}}$ and $U^{-1}_{R,{\rm hf}}$ of the covering groups
$\til{U}(k)$ and $\til{U}(M)$ of $U(k)$ and $U(M)$, respectively, on
which the square-root of the determinant is def\/ined.  A
straightforward exercise leads to the decomposition
\be
\exp(\C^k\ot\C^M) \stackrel{{\rm hf}}{\simeq} \bigoplus_{{\sf l}\in
D_M} \H_{{\sf l}+\half M}^{\til{U}(k)}\ot \overline{\H}_{{\sf l}+\half
k}^{\til{U}(M)} \ll{howehf}
\ee
under $U_{L,{\rm hf}}(\til{U}(k))\ot
U^{-1}_{R,{\rm hf}}(\til{U}(M))$.  Here ${\sf l}+\half M$, regarded as
a highest weight, has components $(l_1+\half M, l_2+\half M,\ldots)$,
and analogously for ${\sf l}+\half k$.  Hence $\H_{{\sf l}+\half M}$
carries the tensor product of the \rep\ of $\til{U}(k)$ characterized
by the Young diagram $\sf l$, and the determinant \rep\ to the power
$M/2$, etc.  This will be further discussed in subsection
\ref{discussion}.

\subsection{Representations induced from $U(M,N)$}
We are now going to attempt to `quantize' Theorem \ref{omw} for $N\neq
0$.  The f\/irst problem is f\/inding a unitary \rep\ of $H=U(M,N)$ that
corresponds to the dominant integral weight $\wmn$ on $\mathfrak t$ (or
the corresponding coadjoint orbit in $\h^*$, cf.\ subsection 2.2);
this is the \rep\ we should induce from.  This problem was solved in
\ci{Ada1}, partly on the basis of the classif\/ication of all unitary
highest-weight modules of $U(M,N)$ \ci{EHW,Jak,Ols2}. In the compact
case, each dominant integral weight corresponds to an irreducible
unitary \rep\ with this weight as its highest weight. For $U(M,N)$ on
the other hand, there are two new phenomena. Firstly, there are
further conditions on the dominant integral weight $\wmn$, namely that
all entries of $\sf m$ should be dif\/ferent, and that all entries of
$\sf n$ should be dif\/ferent. Secondly, the \rep\ corresponding to
$\wmn$, albeit a highest weight \rep, does not in fact have $\wmn$ as
its highest weight. Rather, the highest weight corresponding to $\wmn$
is `renormalized': with $m_1>m_2>\ldots>m_M>0$ and
$n_1>n_2>\ldots>n_N>0$, the highest weight (naively expected to be
$(m_1,\ldots,m_M,-n_N,\ldots,-n_1)$) is in fact
$$
(m_1
+\half(N-M)+\half,\ldots,m_i+\half(N-M)+i-\half,\ldots, m_M+\half
(N+M)-\half,
$$
$$
-(n_N+\half(M+N)-\half),\ldots,-(n_j+\half(M-N)+j-\half),\ldots,
-(n_1+\half(M-N)+\half)).
$$
Note that this highest weight is still
dominant; however, it may no longer be integral, so that it def\/ines a
projective \rep\ of $U(M,N)$ (single-valued on its double cover
$\til{U}(M,N)$).  These highest weight \rep s belong to the
holomorphic discrete series of $U(M,N)$ \ci{Kna}.

The second problem is the quantization of $S=\H\ot\C^{M+N}$, with the
corresponding actions of $G=U_0(\H)$ and $H=U(M,N)$.  One regards
$U(M,N)$ as a subgroup of $Sp(2(M+N),\R)$, so that the symplectic
action of the former on $\C^{M+N}$ is the restriction of the action of
the latter \ci{Ste,KKS}.  Due to the special way we def\/ined the
$U(M,N)$ action in subsection 2.2 as the inverse of a right-action,
the quantization of this action of $Sp(2(M+N),\R)$ is then given by
the conjugate of the metaplectic \rep\ $U_m$ on $L^2(\R^{M+N})\equiv
{\cal L}$, cf.\ \ci{KV,SW,Ste}.  This def\/ines a \rep\ of the inverse
image $\til{U}(M,N)$ of $U(M,N)$ in the metaplectic group
$Mp(2(M+N),\R)$ on $\overline{\cal L}$, which descends to a projective
\rep\ of $U(M,N)$, which we denote by $U_{{\rm hf}}(\til{U}(M,N))$. As
pointed out in \ci{SW} and \ci{BR} (for $k=1$), this \rep\ is
precisely the one obtained from geometric quantization (in a suitable
cohomological variant) if half-forms are taken into account.  This
yields a f\/irst candidate for the quantization of the $U(M,N)$ action
on $\C^{M+N}$.

The second possibility is to take the tensor product of the
(restriction of) the metaplectic \rep\ of $\til{U}(M,N)$ with the
square-root of the determinant, which does def\/ine a unitary \rep\
$U_{{\rm sq}}$ of $U(M,N)$ \ci{SW}; see \ci{BR} for a construction of
this \rep\ from geometric quantization. It is the \rep\ which might be
thought of as being def\/ined by the physicists' second quantization on
$\exp(\C^{M+N})$, as in the $U(M)$ case. However, since the action of
$U(M,N)$ on $\C^{M+N}$ is not unitary, this second quantization is
not, in fact, def\/ined. In geometric quantization this lack of
unitarity shows up through the non-existence of a totally complex
invariant polarization on $S$ which is positive. Consequently, one
needs to work with an indef\/inite such polarization \ci{BR}, and this
leads to complications that will eventually cause a shift in the \rep
s one would naively expect to occur in the decomposition of the
quantization of $S$.

 For f\/inite-dimensional $\H=\C^k$ we therefore have a suitable
quantization of $S=\C^k\ot\C^{M+N}$, namely the Hilbert space
$\Lk\equiv \ot^k\overline{\cal L}$ (the Fock space realization of this
space is not useful, so we drop the notation $\F$). Moreover, we have
natural unitary \rep s $\ot^k U_{{\rm sq/hf}}$ of $\til{U}(M,N)$ on
$\Lk$, which are quantizations of the symplectic action of $U(M,N)$ on
$S$. Following our notation for $U(M)$, we refer to these \rep s as
$U^{-1}_{R,{\rm sq/hf}}$.

 In addition, the quantization of the $U(k)$ action on $S$ may be
found (much more easily) from geometric quantization with or without
half-forms.  The latter case, in which we call the \rep\ $U_{L,{\rm
sq}}(U(k))$, is explicitly given in \ci{KV}.  Its half-form variant
$U_{L,{\rm hf}}(U(k))$ dif\/fers from it by the determinant \rep\ raised
to the power $(M-N)/2$.

 It follows from the theory of Howe dual pairs \ci{How1} that $\Lk$
decomposes discretely under these \rep s.  Starting with $U_{L,{\rm
sq}}(U(k))\ot U^{-1}_{R,{\rm sq}}(U(M,N))$, the explicit decomposition
of $\Lk$ is given in \ci{KV} as (remember that we have to take the
conjugate of the $U(M,N)$ modules, but not of the $U(k)$ modules used
in \ci{KV}, since our $U(k)$ action is the usual one; also, we use the
conventions of \ci{Ada1} and \ci{How2} for labelling the highest
weight, rather than those of \ci{KV} -- this corresponds to an
interchange of $\sf m$ and $\sf n$)
\be
\Lk \stackrel{{\rm sq}}{\simeq} \bigoplus_{\wmn}
\H_{\wmn}^{U(k)}\ot \overline{\H}_{({\sf
m}+ k,{\sf n})}^{U(M,N)}, \ll{kave}
\ee
where the sum is over all
pairs $\wmn$ as def\/ined before, with zeros allowed, but neither $\sf
m$ nor $\sf n$ allowed to be empty. $\H_{\wmn}^{U(k)}$ as a \rep\
space of $U(k)$ was def\/ined in subsection 2.3, and $\H_{({\sf
m}+k,{\sf n})}^{U(M,N)}$ carries the unitary \rep\ of $U(M,N)$ with
highest weight (not subject to further `renormalization')
\[
(m_1+k,\ldots,m_i+k,\ldots,m_M+k,-n_N,\ldots,-n_j,\ldots,-n_1).
\]

The decomposition under $U_{L,{\rm hf}}(U(k))\ot U^{-1}_{R,{\rm
hf}}(U(M,N))$, on the other hand, reads \ci{How2}
\be
\Lk \stackrel{{\rm hf}}{\simeq} \bigoplus_{\wmn} \H_{({\sf
m}+\half(M-N),{\sf n}-\half(M-N))}^{\til{U}(k)}\ot
\overline{\H}_{({\sf m}+ \half k,{\sf n} +\half k)}^{\til{U}(M,N)},
\ll{kave2}
\ee
where the highest weight $({\sf m}+ \half k,{\sf n}
+\half k)$ is explicitly given by
\[
(m_1+k/2,\ldots,m_i+k/2,\ldots,m_M+k/2,-n_N-k/2,
\ldots,n_j-k/2,\ldots,-n_1-k/2),
\]
whereas $\H_{({\sf m}+\half(M-N),{\sf n}-\half(M-N))}$ is the tensor
product of $\H_{\wmn}$, and $\C$, carrying the determinant \rep\ of
$U(k)$ to the power $(M-N)/2$, cf.\ \ci{How2}).

Working with (\ref{kave}) for the sake of concreteness, we now wish to
apply Rief\/fel induction from a suitable \rep\ of $H=U(M,N)$ to $\Lk$
in order to extract the copy of $\H_{\wmn}^{U(k)}$ for the value of
$\wmn$ selected by the \rep\ we induce from.  Firstly, we need a dense
subspace $L\subset \Lk$ such that the function $x\raw ( U^{-1}_{R,{\rm
sq}}(x)\ps,\phv)$ is in $L^1(H)$ for all $\ps,\phv\in L$. This is
easily found: using the decomposition (\ref{kave}), we take $L$ to
consist of vectors having a f\/inite number of components in the
decomposition, each component of which is in the tensor product of
$\H^{U(k)}_{\ldots}$ and the dense subspace of $K$-f\/inite vectors in
the other factor. Since each function of the type $x\raw
(U(x)\ps,\phv)$, where $U$ is in the discrete series, and $\ps$ and
$\phv$ are $K$-f\/inite vectors, is in Harish-Chandra's Schwartz
space~\ci{Kna} (which is a subspace of $L^1(H)$), this choice indeed
satisf\/ies the demand.  (Based on the explicit realization of $\Lk$ as
a function space \ci{KV}, a more `geometric' choice of $L$ may also be
found.)

As we are going to induce from holomorphic discrete series \rep s of
$U(M,N)$, let us examine the tensor product $\overline{\H}_{({\sf
m}_1,{\sf n}_1)}^{U(M,N)}\ot \H_{({\sf m}_2,{\sf n}_2)}^{U(M,N)}$.
Recall that $\wmn$ (which here refers to the actual highest weight,
rather than the dominant integral weight that is subject to
renormalization, as sketched above) def\/ines a unitary irreducible
\rep\ $U_{\wmn}$ of the maximal compact subgroup $K=U(M)\times U(N)$
with highest weight $(m_1,\ldots,m_M,-n_N,\ldots,-n_1)$.  By Theorem 2
in \ci{Rep}, the above tensor product is unita\-ri\-ly equivalent as a
\rep\ space of $U(M,N)$ to the \rep\ induced (in the usual, Mackey,
sense) from $\overline{U}_{({\sf m}_1,{\sf n}_1)}\ot U_{({\sf
m}_2,{\sf n}_2)}(K)$. Using the reduction-induction theorem, we can
therefore decompose this induced \rep\ as a direct sum over the \rep s
induced from the components in the decomposition of
$\overline{U}_{({\sf m}_1,{\sf n}_1)}\ot U_{({\sf m}_2,{\sf
n}_2)}(K)$.

Let us examine a generic \rep\ $U^{\kappa}(H)$ (realized on the
Hilbert space $\H^{\kappa}$ of functions $\ps:G\raw\H_{\kappa}$
satisfying the equivariance condition
$\ps(xk)=U_{\kappa}(k^{-1})\ps(x)$) induced from an irreducible \rep\
$U_{\kappa}(K)$. The Rief\/fel induction procedure produces the
semi-def\/inite form $(\cdot ,\cdot )_0$ on $L\ot\H_{\ch}$ (where, in
this case, $\H_{\ch}=\H_{({\sf m},{\sf n})}^{U(M,N)}$ for certain
$\wmn$). Using (\ref{kave}) and the previous paragraph, we f\/ind that
$L\ot\H_{\ch}$ is a certain dense subspace of a direct sum with
components of the type $\H_{\wmn}^{U(k)}\ot\H^{\kappa}$, in which $H$
acts trivially on the f\/irst factor. By our construction of $L$, each
element of $L\ot\H_{\ch}$ only has components in a f\/inite number of
these Hilbert spaces, so that we can investigate each component
separately. (Had the number of components of elements of $L$ been
inf\/inite, the study of $(\cdot ,\cdot )_0$ would have been more
involved, as this is an unbounded and non-closable quadratic form, so
that $(\sum_i\ps_i,\phv)_0\neq \sum_i(\ps_i,\phv)_0$ for inf\/inite
sums.)

Factorizing $\int_H dx= \int_N dn\,\int_K dk$ \ci{Kna}, it follows
from the equivariance condition and the orthogonality relations for
compact groups that in a given component
$\H_{\wmn}^{U(k)}\ot\H^{\kappa}$ the expression $(\ps,\phv)_0=\int_H
dx\, ({\mathbb I}\ot U^{\kappa}(x)\ps,\phv)$ vanishes unless $U_{\kappa}$
is the identity \rep\ $U_{\rm id}$ of $K$.  Given a highest weight
\rep\ $U_{\ch}(H)$ we Rief\/fel-induce from, there exists a unique pair
$\wmn$ for which $\H_{\wmn}^{U(k)}\ot\H^{\rm id}$ occurs in the
decomposition of $\Lk\ot\H_{\ch}$ as a sum over induced \rep s of $H$
in the above sense.

 Let $L^{\rm id}$ be the projection of $L\ot\H_{\ch}$ onto this
$\H_{\wmn}^{U(k)}\ot\H^{\rm id}$.  We def\/ine $\til{V}:L^{\rm id}\raw
\H_{\wmn}^{U(k)}$ by linear extension of
$\til{V}\ps_1\ot\ps_2=\ps_1\int_Hdx\, \ps_2(x)$ (where $\ps_1\in
\H_{\wmn}^{U(k)}$ and $\ps_2\in\H^{\rm id}\subset L^2(G)$). The
integral exists by our assumptions on $L$; moreover, the explicit form
of the inner product in $\H^{\rm id}$ (namely $(f,g)=\int_H dx\,
f(x)\overline{g(x)}$, as $K$ is compact) leads to the equality
$(\til{V}\ps,\til{V}\phv)=(\ps,\phv)_0$ (where the inner product on
the left-hand side is the one in $\H_{\wmn}^{U(k)}$). We now extend
$\til{V}$ to a map $V$ from $L\ot\H_{\ch}$ to $\H_{\wmn}^{U(k)}$ by
putting it equal to zero on all spaces involving a factor
$\H^{\kappa}$, where $\kappa\neq {\rm id}$ (and equal to $V$ on
$L^{\rm id}$, of course). Clearly, by this and the preceding
paragraph,
\be
(V\ps,V\phv)=(\ps,\phv)_0. \ll{V}
\ee
We are now in a
standard situation in the theory of Riefel induction, in which we can
identify the null space of $(\cdot ,\cdot )_0$ with the kernel of $V$,
and the induced space $\H^{\ch}$ (which, we recall, is the completion
of the quotient of $L\ot\H_{\ch}$ by this null space in the inner
product obtained from this form) with the closure of the image of
$V$. It is clear from our def\/inition of $L$ that the image of $V$
actually coincides with $\H_{\wmn}^{U(k)}$. Also, the def\/inition of
the induced \rep\ $U^{\ch}$ of $G=U(k)$ on $\H^{\ch}$ immediately
implies that $U^{\ch}\simeq U_{\wmn}$. Finally, note that (\ref{V})
shows explicitly that $(\cdot,\cdot)_0$ is positive semi-def\/inite, a
fact which was already certif\/ied by Prop.\ 2 in \ci{NPL93}.

Putting these arguments together, we have proved:
\begin{theorem}\ll{old3}
Let $U(k)$ and $U(M,N)$ act on $S=\C^k\ot\C^{M+N}$ (equipped with the
 symplectic form (\ref{ommn})) from the left and the right,
 respectively, in the natural way, and let $\Lk$ be the quantization
 of $S$, with commuting \rep s of $U(k)$ and $U(M,N)$ on $\Lk$ (which
 quantize the above symplectic actions) as given (up to conjugation of
 the \rep\ of $U(M,N)$) by Kashiwara-Vergne \ci{KV}.

Then Rieffel induction on $\Lk$ from the holomorphic discrete series
\rep\ of $U(M,N)$ with highest weight $({\sf m}+ k,{\sf n})$ (that is,
the highest weight with components
$(m_1+k,\ldots,m_M+k,-n_N,\ldots,-n_1)$) leads to an induced space
$\Hlmn^{U(k)}$, which as a Rieffel-induced $U(k)$ module carries the
\rep\ $U_{\wmn}(U(k))$ (which is the Young product of the \rep\ with
Young diagram $\sf m$ and the conjugate of the \rep\ with Young
diagram $\sf n$).

Moreover, the induced space is empty if one induces from a highest
 weight \rep\ of $U(M,N)$ of the form $\wmn$ in which at least one
 $m_i$ is smaller than $k$, or is not integral. \ll{dis} \end{theorem}

\section{Discussion}
{The last part of Theorem \ref{old3} is particularly unpleasant for
the quantization theory of constrained system, for it shows that
Theorem \ref{omw} cannot really be `quantized' unless $\sf m$ or $\sf
n$ are empty. For we would naturally induce from the holomorphic
discrete series \rep\ of $U(M,N)$ having the `renormalized' highest
weight corresponding to a coadjoint orbit characterized by $\wmn$, as
explained at the beginning of this subsection. But then for $k$ large
enough the induced space will be empty, rather than consisting of
$\Hlmn^{U(k)}$, as desired. As we have seen, the induction procedure
is only successful if we induce from a \rep\ with highest weight
$({\sf m}+k,{\sf n})$, rather than from the ($k$-independent)
renormalized weight we ought to use by f\/irst principles.  This is
bizarre, given that the original weight $\wmn$ (or the orbit it
corresponds to) knows nothing about $k$ or $U(k)$. In addition, even
without this problem the induced space will often be empty, because
the `correct' renormalized highest weight one induces from may simply
not occur in the Kashiwara-Vergne decomposition (\ref{kave}) because
of the half-integral nature of its entries (which is a pure `quantum'
phenomenon).  (In a rather dif\/ferent setting, the discrepancy for
large $k$ between the `decomposition' of $S$ into pairs of matched
coadjoint orbits for $U(k)$ and $U(M,N)$, and the decomposition of
$\Lk$ under these groups, must have been noticed by Adams \ci{Ada2},
who points out that there is a good correspondence for $k\leq {\rm
min}\,(M,N)$ only.)

It is peculiar to the non-compact ($N\neq 0$) case that this
dif\/f\/iculty even arises if the half-form correction to quantization is
not applied.  For (\ref{kave}) is the non-compact analogue of
(\ref{howe}), and in the latter quantization clearly does commute with
reduction. If we do incorporate half-forms, we obtain (\ref{kave2})
for $U(M,N)$ and (\ref{howehf}) for $U(M)$. In both cases the Rief\/fel
induction process generically (that is, if $M\neq N$) fails to produce
the correct \rep\ of $U(k)$, even if one induces from a \rep\ whose
highest weight is renormalized (compared to the weight expected from
the orbit correspondence) by the term $k/2$.

 Finally, the passage from $\C^k$ to inf\/inite-dimensional Hilbert
spaces is tortuous whenever half-forms are used (the corrections being
inf\/inite for $k=\infty$), and in the non-compact case even without
these.  This is partly because of the $k$-dependence of the highest
weights of $U(M,N)$, and partly because $\cal L$ does not contain the
identity \rep\ of $U(M,N)$ (recall that in the compact case we used
the carrier space $\C\Omega$ of this \rep\ as the f\/ixed `tail' vector
to construct the von Neumann inf\/inite tensor product from).

Clearly, this situation deserves further study. We do not think it is
an artifact of our proposal of using Rief\/fel induction in the
quantization of constrained systems. In fact, this technique comprises
the only method known to us which is precise enough to bring the
embarrassment to light.  \ll{discussion} }

\label{landsman-lp}
\end{document}